\documentclass[10pt,letterpaper]{article}

\usepackage{cvpr}
\usepackage{times}
\usepackage{epsfig}
\usepackage{graphicx}
\usepackage{amsmath}
\usepackage{amssymb}

\usepackage{lipsum}
\usepackage{subcaption}

\usepackage[pagebackref=true,breaklinks=true,letterpaper=true,colorlinks,bookmarks=false]{hyperref}

\cvprfinalcopy 

\def\Ttheta{\boldsymbol\theta}

\def\TC{\mathbf C}

\def\TK{\mathbf K}

\def\TY{\mathbf Y}

\def\TW{\mathbf W}

\def\TX{\mathbf X}

\ifcvprfinal\fi
\begin{document}

\title{Fully reversible neural networks for large-scale surface and sub-surface characterization via remote sensing}

\author{Bas Peters\\
Computational Geosciences Inc.\\
1623 West 2nd Ave, Vancouver, BC, Canada\\
{\tt\small bas@compgeoinc.com}
\and
Eldad Haber\\
University of British Columbia\\
2207 Main Mall, Vancouver, BC, Canada\\
{\tt\small ehaber@eoas.ubc.ca}
\and
Keegan Lensink\\
University of British Columbia\\
2207 Main Mall, Vancouver, BC, Canada\\
{\tt\small klensink.ubc.ca}
}

\maketitle

\begin{abstract}
The large spatial/frequency scale of hyperspectral and airborne magnetic and gravitational data causes memory issues when using convolutional neural networks for (sub-) surface characterization. Recently developed fully reversible networks can mostly avoid memory limitations by virtue of having a low and fixed memory requirement for storing network states, as opposed to the typical linear memory growth with depth. Fully reversible networks enable the training of deep neural networks that take in entire data volumes, and create semantic segmentations in one go. This approach avoids the need to work in small patches or map a data patch to the class of just the central pixel. The cross-entropy loss function requires small modifications to work in conjunction with a fully reversible network and learn from sparsely sampled labels without ever seeing fully labeled ground truth. We show examples from land-use change detection from hyperspectral time-lapse data, and regional aquifer mapping from airborne geophysical and geological data.
\end{abstract}

\section{Introduction}
Remote sensing datasets collected by satellites and aircraft are relatively convenient to work with because they cover large areas, we can easily cut out rectangular patches, and connect well with computer vision techniques including convolutional neural networks (CNNs). However, many CNNs assume there are hundreds or more images with often just 3-channel input. Most benchmarks come with fully annotated/segmented/labeled images. The reality in remote sensing is different: large images and hundreds of hyperspectral frequencies as input. While there may be an abundance of hyperspectral data itself, this does not hold for labels. Many land-use classification applications assume some sparse ground truth and aim to interpolate/extrapolate using hyperspectral data, see, e.g., \cite{C7729859,M8297014,doi:10.1080/2150704X.2019.1681598,rs9010067,rs12010188}. The observations above about hyperspectral data and labels also hold for shallow sub-surface remote sensing based on gravitational, magnetic, topographical, and geological data. 

In this work, we discuss the similarities and the subtle differences between multi-modality sensing for sub-surface applications, and hyperspectral sensing for surface characterization. Both types of remote sensing often come with large-scale inputs in terms of spatial size and number of frequencies and modalities. This causes memory issues with networks that were developed for much smaller images with RGB inputs. A second challenge arises when there is only sparse ground truth available for training. Many works on hyperspectral land-use classification approach the problem by training networks to classify the central pixel of a small patch of data \cite{K7326945,M8297014,L8851917,rs9010067}. This approach allows sparse sampling of the labels and reduces the size of the input to mitigate memory limitations on graphical processing units (GPUs). However, it prevents the network from having access to the larger (spatial) structures present in data.

Our primary contributions are threefold:
\begin{itemize}
\item This is the first work, to the best of our knowledge, where fully reversible neural networks for semantic segmentation \cite{lensink2019fully,peters2019symmetric} enable learning from remote sensing data on a much larger scale than before while also working with arbitrarily deep networks. These benefits result from the fact that fully reversible networks have memory requirements that are independent of network depth and number of pooling layers.

\item We make small modifications to the standard semantic segmentation problem setup and cross-entropy loss function so that fully reversible networks can learn from sparsely sampled ground truth.

\item We present examples that show there are subtle differences between learning from hyperspectral data, and multimodality data acquired by aircraft, including gravitational and magnetic data. These small differences do not prevent us to use the developed tools to solve both problems in almost the same way.
\end{itemize}

After reviewing fully reversible convolutional neural networks, we show how to train both on hyperspectral and multi-modality datasets with sparse spatial label sampling. This discussion includes the slight differences between the two scenarios considered. Examples include land-use change detection from time-lapse hyperspectral data and sub-surface aquifer mapping.

\section{Fully reversible neural networks for large scale remote sensing}

Most neural networks for classification and segmentation from remote sensing data use convolutional kernels because of their performance on vision tasks and relatively low parameter count per network layer. The literature contains little discussion about the computation of the gradients of the loss functions. This is because the overwhelming majority relies on reverse-mode automatic differentiation. This type of gradient computation requires access to the network states (activations) $\TY_j$ at layer $j$ during the backpropagation phase. Standard implementations keep all network states in memory, causing the memory footprint to grow linearly with network depth. Workarounds often rely on relatively shallow networks ($\approx \leq 9$ layers) or a network design that maps a small patch or data sub-volume into the class of the central pixel/voxel. 

Because fully reversible networks require memory for the states of just three layers, there is no longer a need to trade-off depth for input size. The memory savings by using reversible architectures allow us to allocate all available memory towards larger data input volumes, which enables the network to learn from large-scale structures.

Various reversible networks were proposed for image classification \cite{DinhSB16,Chang2017Reversible,GomezEtAl2017}. We refer to such networks as block-reversible because they are reversible in between pooling/coarsening operations and require storing additional network states as checkpoints before each pooling layer. Fully reversible networks that contain reversible or invertible coarsening operations were proposed for image classification \cite{ jacobsen2018irevnet,leemput2018memcnn} and image/video segmentation \cite{lensink2019fully,peters2019symmetric}. The latter uses the orthogonal Haar wavelet transform $\TW$ to coarsen the image and increase the number of channels. The transpose achieves the reverse of these operations, i.e., the action of the linear operator $\TW$ on a tensor $\TY$ creates the mappings
\begin{align}
\TW \TY &: \mathbb{R}^{n_1 \times n_2 \times n_3 \times n_\text{chan}} \rightarrow \mathbb{R}^{n_1/2 \times n_2/2 \times n_3/2 \times 8n_\text{chan}}, \\
\TW^{-1} \TY&: \mathbb{R}^{n_1 \times n_2 \times n_3 \times n_\text{chan}} \rightarrow \mathbb{R}^{2 n_1 \times 2 n_2 \times 2 n_3 \times n_\text{chan}/8}.
\end{align}
Because of the invertibility of any orthogonal transform, applying $\TW$ and $\TW^\top$ incurs no loss of information.

A conservative leapfrog discretization of non-linear Telegraph equation with time-step $h$ is the basis for the reversible architecture of \cite{Chang2017Reversible}. Combined with the orthogonal wavelet transform, $\TW$, for changing resolution and the number of channels \cite{lensink2019fully}, the network recursion reads
\begin{align}\label{network}
\TY_1 & = \TX, \quad \TY_2 = \TX \\
\TY_{j} &= 2 \TW_{j-1}\TY_{j-1} -  \TW_{j-2} \TY_{j-2} 
-  h^2 \TK(\Ttheta_{j})^\top f( \TK(\Ttheta_{j}) \TW_{j-1} \TY_{j-1}), 
 \text{for} \: j=3,\cdots,n. \nonumber
\end{align}
The first to states are the initial conditions, which we set equal to the input data $\TX \in \mathbb{R}^{n_1 \times n_2 \times n_2 \times n_3 \times n_\text{chan}}$. Note that we set $\TW_j$ as the identity if we do not want to change resolution at layer $j$. The `time-step' $h$ affects the stability of the forward propagation \cite{HaberRuthotto2017a}. The linear operator $\TK(\Ttheta_j)$ is a representation of the convolutions with kernels $\Ttheta_j$. In this work, we select the ReLU as the pointwise non-linear activation function $f$. 

Reversibility of the above relation is the key property. By isolating one of the states in \eqref{network} and shifting indices, we obtain an expression for the current state in terms of future states:
\begin{equation}\label{rev_prop}
\TY_j = \TW_j^{-1} \bigg[ 2 \TW_{j+1} \TY_{j+1} 
-  h^2 \TK(\Ttheta_{j+2})^\top  f ( \TK(\Ttheta_{j+2}) \TW_{j+1} \TY_{j+1} ) - \TY_{j+2} \bigg] 
 \text{for} \: j=n-2,\cdots,3.
\end{equation}
This equation does not require inverting the activation function $f$. Instead, only the inversion of the orthogonal wavelet transform is required, which is known in closed form. When computing the gradient of the loss function using backpropagation, we recompute the states $\TY_j$ while going back through the network. The recomputation avoids the storage of all $\TY_j$ and leads to a fixed memory requirement for the states of three layers, see Table \ref{memory} for an overview.

\begin{table*}[]
\caption{Memory requirements for the states $\TY_j$ for fully reversible and non-reversible equivalent networks based on the networks in Table \ref{network_design}.}
\label{memory}
\begin{center}
\begin{tabular}{lclcl}
\multicolumn{1}{c}{Memory (megabyte)}  &\multicolumn{1}{c}{\bf Fully reversible network} &\multicolumn{1}{c}{\bf Non-reversible}
\\ \hline \\
Rate         & $3 n_1 n_2 n_3 n_\text{chan}$  & $n_\text{layers} n_1 n_2 n_3 n_\text{chan}$\\
Hyperspectral example             & 798 & $5057$\\
Aquifer mapping example           & 1268 & $6763$
\end{tabular}
\end{center}
\end{table*}

\section{Spatial semantic segmentations from 3D and 4D data using fully reversible networks}

The goal is to create a spatial map from 3D/4D hyperspectral or other remote sensing data. Although relatively standard, we cannot straightforwardly use the cross-entropy loss function because fully reversible networks output a tensor of the same size as the input by construction. However, we are just interested in a spatial map of the earth in terms of a semantic segmentation of land-use or aquifer presence. 

Non-reversible networks can output a tensor that has a size different from the input by pooling/coarsening in one direction more than another, or by reducing the number of channels. Both of these approaches can `compress' a 3D or 4D tensor into a 2D matrix.

Fully reversible networks require a different approach. We propose to embed the known ground-truth labels in the label tensor $\TC \in \mathbb{R}^{n_1 \times n_2 \times n_3 \times n_\text{chan}}$ at slice $p$ as $\TC_{:,:,p,1:n_\text{class}}$. The number of different classes, $n_\text{class}$, has to be smaller or equal to $n_\text{chan}$. All other entries in the label tensor are unknown and not used in the loss and do not contribute to the gradient computation. The resulting multi-class cross entropy function with softmax reads
\begin{equation}\label{partial_ce}
l(\TX,\Ttheta,\TC) = 
- \sum_{(i,j) \in \Omega}  \sum_{k=1}^{n_\text{class}} \TC_{i,j,p,k} \log \bigg( \frac{\operatorname{exp} (g(\Ttheta,\TX)_{i,j,p,k})}{\sum_{k=1}^{n_\text{class}}\operatorname{exp}(g(\Ttheta,\TX)_{i,j,p,k})} \bigg),
\end{equation}
where the spatial location indices of known labels are collected in the set $\Omega$. The nonlinear function $g(\Ttheta,\TX)$ denotes a neural network for which the inputs are parameters $\Ttheta$ and data $\TX$. The loss function is thus a sum over sparse spatial locations in a single slice slice ($p$) and over a part of the channels.  Computationally, the loss function requires a full forward pass trough the network, followed by sampling at the indices of interest that correspond to known labels locations. All entries where the labels are not known or that are not part of our problem formulation do not contribute to the subsequent gradient computation. See \cite{doi:10.1190/INT-2018-0225.1,doi:10.1190/tle38070534.1} for more information and applications of partial loss functions.

This section showed that two small modifications to a standard cross-entropy loss enable us to \emph{a}) learn a mapping from data to a full segmentation while never having access to fully annotated examples; \emph{b}) by embedding map-type labels in a 3D/4D tensor, fully reversible networks also apply to problems where we want to reduce multi-dimensional data types into a 2D spatial map.

\begin{figure}[!htb]
 	\centering
 	\begin{subfigure}[b]{0.4\textwidth}
 		\includegraphics[width=\textwidth]{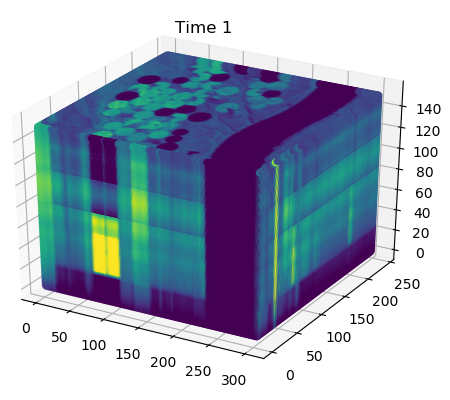}
 		\caption{}
 		\label{fig:Figure1a}
 	\end{subfigure}
 	\begin{subfigure}[b]{0.4\textwidth}
 		\includegraphics[width=\textwidth]{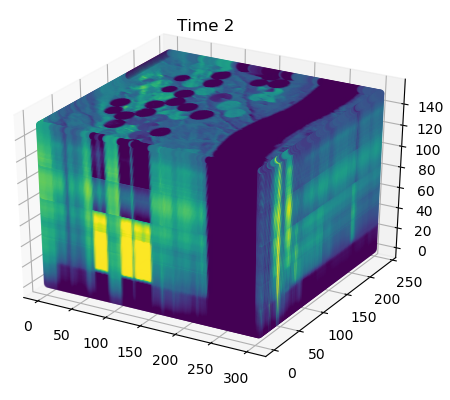}
 		\caption{}
 		\label{fig:Figure1b}
 	\end{subfigure}
 	\caption{The hyperspectral data collected at two different times.}
\label{fig:data}
 \end{figure}
 
 \begin{table*}[]
\caption{Network design for the fully reversible networks. The time-lapse hyperspectral example uses $3\times3\times3$ convolutional kernels. The the number of input channels is increased up from two to $384$ by replicating the data three times and applying two subsequent 3D Haar transforms. The network design for the aquifer mapping example uses $3 \times 3$ convolutional kernels.}
\label{network_design}
\begin{center}
\begin{tabular}{lcl|lcll}
\multicolumn{1}{c}{}  &\multicolumn{1}{c}{\bf Hyperspectral} &\multicolumn{1}{c}{} &\multicolumn{1}{c}{}  &\multicolumn{1}{c}{\bf Aquifer mapping} &\multicolumn{1}{c}{}\\ \\
\multicolumn{1}{c}{\bf Layer}  &\multicolumn{1}{c}{\bf Channels} &\multicolumn{1}{c}{\bf Feature size} &\multicolumn{1}{c}{\bf Layer}  &\multicolumn{1}{c}{\bf Channels} &\multicolumn{1}{c}{\bf Feature size}
\\ \hline \\
1-5         & 384 & $76 \times 60 \times 38$ &  3-4         & 56 & $1296 \times 1456$\\
6-11             & 48 & $152 \times 120 \times 76$ & 5-9             & 224 & $648 \times 728 $\\
12-19             & 6 & $304 \times 240 \times 152$ & 10-18             & 56 & $1296 \times 1456 $
\end{tabular}
\end{center}
\end{table*}

\section{Time-lapse hyperspectral land-use change detection}

\begin{figure}[]
 	\centering
 	\begin{subfigure}[b]{0.22\textwidth}
 		\includegraphics[width=\textwidth]{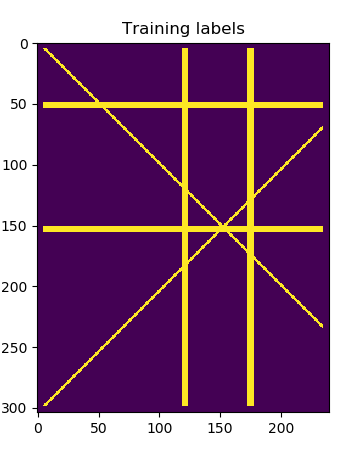}
 		\caption{}
 		\label{fig:Figure1a}
 	\end{subfigure}
 	\begin{subfigure}[b]{0.22\textwidth}
 		\includegraphics[width=\textwidth]{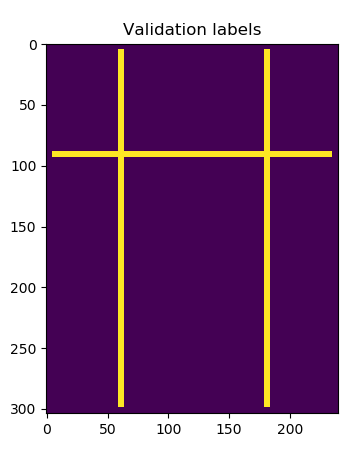}
 		\caption{}
 		\label{fig:Figure1b}
 	\end{subfigure}
 	\caption{Plan view of the label locations for training and validation for the hyperspectral example.}
\label{fig:labels}
 \end{figure}

The data, $\TX$, has two spatial coordinates $n_1$ and $n_2$, and the third dimension corresponds to frequency. There is one channel per time of data collection, two in this example. Figure \ref{fig:data} displays these data sets \cite{doi:10.1080/01431161.2018.1466079}. We follow common practice in hyperspectral imaging literature, where part of the segmentation is assumed known. The lines in Figure \ref{fig:labels} show where there are training and validation labels. The training labels amount for about $9\%$ of the surface. 
    \begin{figure*}[]
   \centering
   \includegraphics[width=0.8\textwidth]{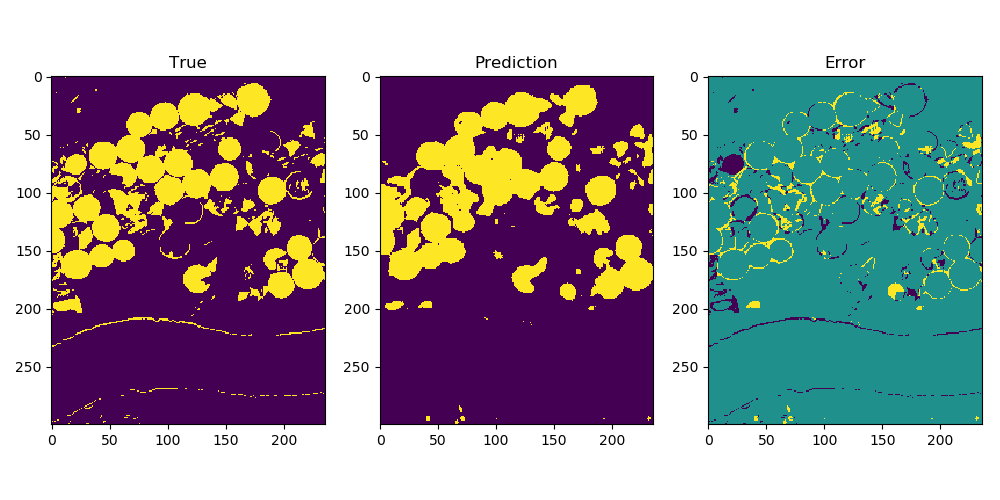}
   \caption{True land-use change, prediction, and the error.}
   \label{fig:prediction_threshold}
 \end{figure*}
 
Table \ref{network_design} contains network details. For training, we decrease the loss \eqref{partial_ce} using stochastic gradient descent with momentum and a decaying learning rate for $320$ iterations. At every iteration, the loss and gradient computation use one-fifth of the known labels shown in Figure \ref{fig:labels}, randomly selected. We also apply random permutations and flips to the two spatial data dimensions. 
 
Figure \ref{fig:prediction_threshold} displays the true land-use change, prediction, and errors. Aside from some boundary artifacts, there are only two center-pivot fields classified incorrectly.
 
\section{Regional-scale aquifer mapping}

\begin{figure*}[b]
\begin{center}
   \includegraphics[width=0.8\textwidth]{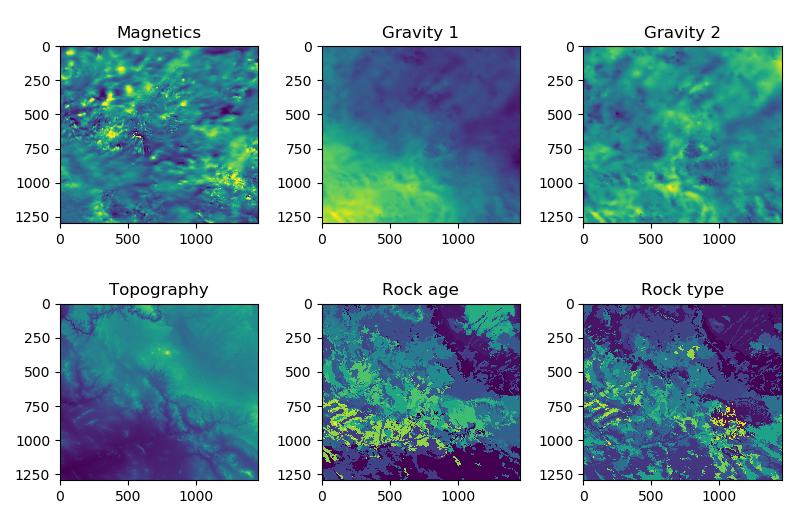}
   \caption{The data inputs for the aquifer mapping example. Each type is placed in a separate channel of the input. We do not use the two geological maps as images. Instead, each class is converted to a map with zero/one values, resulting in 52 separate geological maps. }
   \label{fig:aq_data}
   \end{center}
\end{figure*}

\begin{figure*}[]
\begin{center}
   \includegraphics[width=0.99\textwidth]{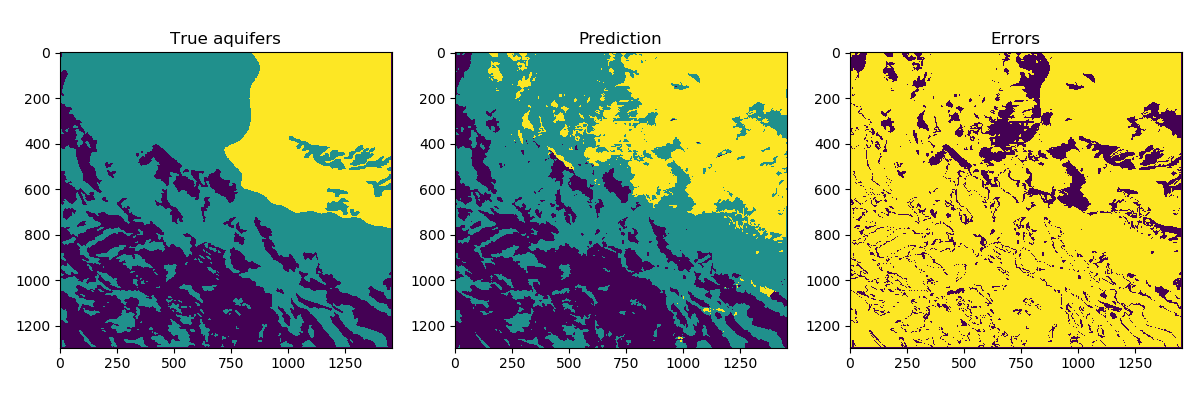}
   \caption{True aquifer map, prediction, and errors in blue, correct in yellow}
   \label{fig:aq_results}
   \end{center}
\end{figure*}

The task is the delineation of large aquifers in Arizona, USA, see Figure \ref{fig:aq_results}. The classes are: basin and range aquifers, Colorado Plateau aquifer, and no aquifer \cite{wateratlas}. The survey area is almost the entire state. Aircraft-based sensors collected magnetic and two types of gravity measurements, see Figure \ref{fig:aq_data}. 

We also use the topography. Besides these remotely acquired data, we supplement two types of geological maps: one map in terms of rock age, and one in terms of rock types. The advantage of using geological maps is that they incorporate expert knowledge into our data. Geologists construct these maps by synthesizing their geological knowledge with ground truth observations, hyperspectral data, and various airborne and land-based geophysical surveys. A disadvantage of using geological maps is that their resolution is unknown because they are partly composed of other geological maps created on various scales. The geological maps in Figure \ref{fig:aq_data} are not invariant under the permutation of the class numbers. This would influence what the network will `see'. Therefore, we create one map per class that shows where a particular rock class is present or not, resulting in 52 separate geological maps derived from the two original maps. 

\begin{figure}[t]
\begin{center}
   \includegraphics[width=0.49\textwidth]{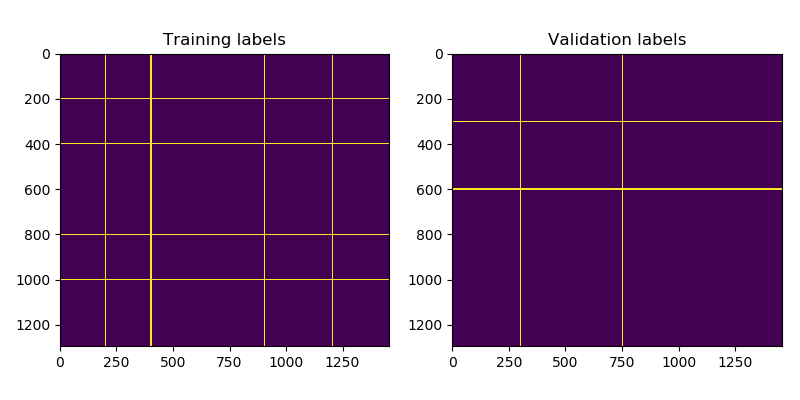}
   \caption{Locations of the training and validation labels for the aquifer mapping example.}
   \label{fig:aq_labels}
   \end{center}
\end{figure}

Just like in the hyperspectral example, we assume partial ground truth as if an expert annotated the aquifers along a few transects, see Figure \ref{fig:aq_labels}. Training is similar to the previous example: SGD+momentum with a decaying learning rate for $140$ iterations. Each iteration uses about $1/5$ of the known labels to compute an approximation of the loss and the gradient. We also augment the data with random flips and permutations. The network details can be found in Table \ref{network_design}. Figure \ref{fig:aq_results} displays the results and errors. Most of the error concentrates on a few patches, as well as minor errors along some of the geological rock type boundaries.

\section{Conclusions}
We presented computational methods for designing and training convolutional neural networks for characterizing both the surface and sub-surface from remote sensing data. Hyperspectral data, airborne geophysical data, as well as geological maps, lead to spatially large images with dozens to hundreds of frequencies or channels. Under tight memory constraints, most neural network approaches are limited to shallow networks and approaches that map from a small patch/subvolume to the class of the central pixel. We showed that the fundamentally lower memory requirements of fully reversible networks enable the semantic segmentation of large 3D and 4D datasets in one go, without resorting to small patches. Minor modifications of standard cross-entropy losses suffice to train in a randomized fashion on a single example with partial ground truth, and without ever having access to fully annotated examples. 
Because fully reversible neural networks were recently developed, this is the first effort to apply them to remote sensing applications. The presented computational tools enable learning on a larger scale and alleviate memory limitations associated with deep networks. The results for land-use change detection and aquifer mapping are encouraging. To evaluate the full benefits, we need to research how to annotate or sample ground truth for label generation from the deep network point of view.

{\small
\bibliographystyle{ieee_fullname}
\bibliography{biblio,VideoRefs,HyperSpectralRefs}

\begin{thebibliography}{10}\itemsep=-1pt

\bibitem{Chang2017Reversible}
Bo Chang, Lili Meng, Eldad Haber, Lars Ruthotto, David Begert, and Elliot
  Holtham.
\newblock Reversible architectures for arbitrarily deep residual neural
  networks.
\newblock In {\em AAAI Conference on AI}, 2018.

\bibitem{DinhSB16}
Laurent Dinh, Jascha Sohl{-}Dickstein, and Samy Bengio.
\newblock Density estimation using real {NVP}.
\newblock {\em CoRR}, abs/1605.08803, 2016.

\bibitem{GomezEtAl2017}
Aidan~N Gomez, Mengye Ren, Raquel Urtasun, and Roger~B Grosse.
\newblock The reversible residual network: Backpropagation without storing
  activations.
\newblock In {\em Adv Neural Inf Process Syst}, pages 2211--2221, 2017.

\bibitem{HaberRuthotto2017a}
Eldad Haber and Lars Ruthotto.
\newblock Stable architectures for deep neural networks.
\newblock {\em Inverse Problems}, 34(1):014004, dec 2017.

\bibitem{doi:10.1080/01431161.2018.1466079}
Mahdi Hasanlou and Seyd~Teymoor Seydi.
\newblock Hyperspectral change detection: an experimental comparative study.
\newblock {\em International Journal of Remote Sensing}, 39(20):7029--7083,
  2018.

\bibitem{M8297014}
M. {He}, B. {Li}, and H. {Chen}.
\newblock Multi-scale 3d deep convolutional neural network for hyperspectral
  image classification.
\newblock In {\em 2017 IEEE International Conference on Image Processing
  (ICIP)}, pages 3904--3908, Sep. 2017.

\bibitem{jacobsen2018irevnet}
J\"{o}rn-Henrik Jacobsen, Arnold~W.M. Smeulders, and Edouard Oyallon.
\newblock i-revnet: Deep invertible networks.
\newblock In {\em International Conference on Learning Representations}, 2018.

\bibitem{C7729859}
H. {Lee} and H. {Kwon}.
\newblock Contextual deep cnn based hyperspectral classification.
\newblock In {\em 2016 IEEE International Geoscience and Remote Sensing
  Symposium (IGARSS)}, pages 3322--3325, July 2016.

\bibitem{lensink2019fully}
Keegan Lensink, Eldad Haber, and Bas Peters.
\newblock Fully hyperbolic convolutional neural networks.
\newblock {\em arXiv preprint arXiv:1905.10484}, 2019.

\bibitem{L8851917}
A. {Li} and Z. {Shang}.
\newblock A new spectral-spatial pseudo-3d dense network for hyperspectral
  image classification.
\newblock In {\em 2019 International Joint Conference on Neural Networks
  (IJCNN)}, pages 1--7, July 2019.

\bibitem{rs9010067}
Ying Li, Haokui Zhang, and Qiang Shen.
\newblock Spectral–spatial classification of hyperspectral imagery with 3d
  convolutional neural network.
\newblock {\em Remote Sensing}, 9(1), 2017.

\bibitem{K7326945}
K. {Makantasis}, K. {Karantzalos}, A. {Doulamis}, and N. {Doulamis}.
\newblock Deep supervised learning for hyperspectral data classification
  through convolutional neural networks.
\newblock In {\em 2015 IEEE International Geoscience and Remote Sensing
  Symposium (IGARSS)}, pages 4959--4962, July 2015.

\bibitem{doi:10.1190/INT-2018-0225.1}
Bas Peters, Justin Granek, and Eldad Haber.
\newblock Multiresolution neural networks for tracking seismic horizons from
  few training images.
\newblock {\em Interpretation}, 7(3):SE201--SE213, 2019.

\bibitem{doi:10.1190/tle38070534.1}
Bas Peters, Eldad Haber, and Justin Granek.
\newblock Neural networks for geophysicists and their application to seismic
  data interpretation.
\newblock {\em The Leading Edge}, 38(7):534--540, 2019.

\bibitem{peters2019symmetric}
Bas Peters, Eldad Haber, and Keegan Lensink.
\newblock Symmetric block-low-rank layers for fully reversible multilevel
  neural networks.
\newblock {\em arXiv preprint arXiv:1912.12137}, 2019.

\bibitem{wateratlas}
Stanley~G. Robson and Edward~R. Banta.
\newblock Ground water atlas of the united states: Segment 2, arizona,
  colorado, new mexico, utah.
\newblock Technical report, U.S. Geological Survey, 1995.

\bibitem{leemput2018memcnn}
Sil~C van~de Leemput, Jonas Teuwen, and Rashindra Manniesing.
\newblock Memcnn: a framework for developing memory efficient deep invertible
  networks.
\newblock 2018.

\bibitem{rs12010188}
Qin Xu, Yong Xiao, Dongyue Wang, and Bin Luo.
\newblock Csa-mso3dcnn: Multiscale octave 3d cnn with channel and spatial
  attention for hyperspectral image classification.
\newblock {\em Remote Sensing}, 12(1), 2020.

\bibitem{doi:10.1080/2150704X.2019.1681598}
Zhixiang Xue.
\newblock A general generative adversarial capsule network for hyperspectral
  image spectral-spatial classification.
\newblock {\em Remote Sensing Letters}, 11(1):19--28, 2020.

\end{thebibliography}
}

\end{document}